\documentclass[aps,prb,twocolumn,groupedaddress]{revtex4-1}
\usepackage{graphicx}
\usepackage{amsmath}
\usepackage{epstopdf}
\begin{document}

\title{Dynamic rotor mode in antiferromagnetic nanoparticles}

\author{K. Lefmann}
\email[]{lefmann@fys.ku.dk}
\author{H. Jacobsen}
\author{J. Garde}
\author{P. Hedeg\aa rd}
\affiliation{Nanoscience Center, Niels Bohr Institute, University of Copenhagen, 2100 Copenhagen {\O}, Denmark}
\author{A. Wischnewski}
\affiliation{Research Center J\"ulich, J\" ulich Center for Neutron Science, D-52425 J\"ulich, Germany}
\author{S. N. Ancona}
\author{H.S. Jacobsen}
\author{C. R. H. Bahl}
\author{L. Theil Kuhn}
\affiliation{Materials Science Division, Ris\o\ DTU, Technical University of Denmark, 4000 Roskilde, Denmark}
\affiliation{Department of Energy Conversion and Storage, Technical University of Denmark, 4000 Roskilde, Denmark}

\date{\today}

\begin{abstract}
We present experimental, numerical, and theoretical evidence for a new mode of
antiferromagnetic dynamics in nanoparticles.
Elastic neutron scattering experiments on 8 nm particles of hematite display
a loss of diffraction intensity with temperature, the intensity vanishing around 150~K.
However, the signal from inelastic neutron scattering remains above that temperature,
indicating a magnetic system in constant motion.
In addition, the precession frequency of the inelastic magnetic signal shows an increase above 100~K.
Numerical Langevin simulations of spin dynamics reproduce all measured neutron data
and reveal that thermally activated spin canting gives rise to a new type of
coherent magnetic precession mode.
This ''rotor`` mode can be seen as a high-temperature version of superparamagnetism
and is driven by exchange interactions between the two magnetic sublattices.
The frequency of the rotor mode behaves in fair agreement with a simple analytical model, 
based on a high temperature approximation of the generally accepted Hamiltonian of the system.
The extracted model parameters, as the magnetic interaction and the axial anisotropy, are in excellent agreement with 
results from M\"ossbauer spectroscopy.
\end{abstract}


\maketitle

\section{Introduction}
Magnetic nanoparticles have important applications in modern technology,
including magnetic data storage, ferrofluids, magnetic resonance imaging,
biotechnology, and biomedicine \cite{lu07, morup08}.
In addition, the study of nanoparticle magnetism have elucidated many fundamental scientific phenomena,
like thermally induced fluctuations, leading to magnetization reversal, known as superparamagnetism (SPM).
The basic theory for SPM was derived
by N\'eel and Brown \cite{neel49,brown63}, who found that the SPM relaxation time
is given by an Arrhenius-like expression,
\begin{equation} \label{eq:neelbrown}
\tau = \tau_0 \exp\left( \frac{K V}{k_{\rm B}T} \right) ,
\end{equation}
where $\tau_0$ is a typical attempt frequency 
and $KV$ is the magnetic anisotropy barrier.
Eq.~(\ref{eq:neelbrown}) is supported by numerous experimental studies, see {\rm e.g.} Refs.~\onlinecite{lu07,morup08}.
Below temperatures where the SPM relaxation is important,
the magnetic dynamics is dominated by a coherent uniform precession mode,
which can be considered as a $q=0$ spin wave.
This mode gives rise to a linear decrease of the magnetization with increasing temperature \cite{morup76}.

The magnetic dynamics of antiferromagnetic (AFM) nanoparticles has attracted much attention,
because it displays much richer intrinsic dynamics than that of ferromagnetic nanoparticles \cite{morup07},
and because complications due to dipolar interactions between particles can be neglected.
In AFM nanoparticles with uniaxial anisotropy, 
the uniform precession mode resembles a $q=0$ AFM spin wave,
and its frequency is influenced both by the anisotropy and by
the strength of the magnetic exchange interaction \cite{bahl08}.

The technique of inelastic neutron scattering have earlier been found to be efficient
to investigate the magnetic dynamics of nanoparticles, 
see {\em e.g.} Refs.~\onlinecite{hennion94,hansen97,lefmann01,klausen03,klausen04,kuhn06,bahl06,feygenson11,disch14,hill14,brok14}.
In particular, in 15~nm hematite ($\alpha$-Fe$_2$O$_3$) particles,
the frequency of the uniform modes was found 
to decrease with increasing temperature. This was ascribed to anharmonicities 
in the uniaxial anisotropy potential \cite{hansen97}.

In this work, we present neutron diffraction data on 8~nm hematite particles,
which show an unexpected vanishing of static magnetic order at temperatures as low as 150~K, despite
the fact that the N\'eel temperature of bulk hematite exceeds 900~K.
We analyse this result in the light of inelastic neutron data,
which show a clear signal from the uniform magnetic mode at all temperatures studied (up to 300~K)
and an increase with temperature of the uniform mode frequency.
We further investigate the physics of this system by a theoretical model 
and by numerical simulations, and we find that they lead towards
the existence of a novel dynamic ''rotor`` mode in AFM nanoparticles.

\section{The nanoparticle sample}
Hematite is a common AFM mineral occuring frequently 
in soils as nanoparticles.
The magnetic ions in hematite are Fe$^{3+}$ ($s=5/2$),
which are essentially arranged in a hexagonal structure. The spins are ferromagnetically
ordered in the $(a,b)$ planes, with alternating directions along the $c$ direction.
In bulk hematite above the Morin transition, $T_{\rm M}=263$~K, 
the spins are confined to the plane
by a strong planar anisotropy, while below $T_{\rm M}$,
the spins turn to point along the $c$-axis \cite{morrish}.

In particles smaller than $\sim 20$~nm, the Morin transition is suppressed.
At low temperatures, the spins align along an easy direction
within the $(a,b)$ plane due to a weak axial anisotropy, presumably from the particle surface.
The sublattices display a small canting from antiparallel within the plane
(in bulk \mbox{$\sim$0.1$^\circ$}\ \cite{morrish}), due to the Dzyaloshinsky-Moria interaction.
Our simulations show that this small term has little effect on the properties we investigate,
and it is generally irrelevant in the following discussion. 
A later publication will discuss this approximation in more detail \cite{jacobsen14}.

The nanoparticle sample of hematite was characterized earlier \cite{kuhn06}. 
The particles are phosphate coated at the surface.
Electron microscopy and X-ray diffraction show that the particles have
an average diameter of 8~nm. The phosphate layer is very thin, most likely a monolayer, 
but sufficient to minimize magnetic interparticle interactions \cite{kuhn06}.

\section{Neutron scattering experiments}
Neutron diffraction experiment were performed at the Paul Scherrer Institute (CH), using the two-axis
powder diffractometer DMC and the triple-axis spectrometer \mbox{RITA-2}. The DMC experiments were performed
with a wavelength of 4.2~\AA\ (4.7~meV), giving an effective energy resolution wider than 5~meV (FWHM),
while the RITA-2 experiments were performed at 4.7~\AA\ (3.7~meV) with an energy resolution of 0.12~meV (FWHM).

The inelastic neutron scattering experiments were performed 
with \mbox{RITA-2} in the monochromatic imaging mode \cite{bahlrita1,bahlrita2}.
These experiments were performed with a constant final energy
of $3.7$~meV, also here with energy resolution of 0.12~meV (FWHM). 

Quasielastic neutron experiments were performed at the backscattering spectrometer BSS at
Research Center J\"ulich, using a neutron wavelength of 6.271~\AA\ and an energy resolution of 1.0~$\mu$eV (FWHM).
Of the 14 detectors of BSS, we used number 2 to 5, corresponding to $q$-values of
1.9, 1.7, 1.5, and 1.3~\AA$^{-1}$, respectively. Each detector covers a $q$-range of around 0.2~\AA$^{-1}$.
Detectors 3, 4, and 5 correspond to the structural ($10\bar{2})$ peak and the magnetic $(101)$ and $(003)$ peaks,
respectively.

\section{Neutron scattering data}
\begin{figure}[]
        \includegraphics[width=0.40\textwidth]{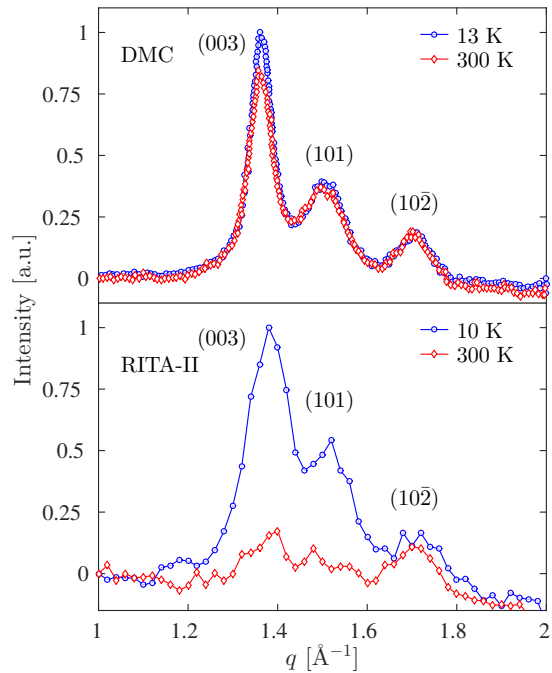}
 \caption{
        (color online) Neutron diffraction data around the magnetic (003) and (101) AFM Bragg peaks,
        corresponding to $q=1.37$~\AA$^{-1}$ and $q=1.51$~\AA$^{-1}$, respectively, taken at 10~K and 300~K.
        (top) data from the two-axis powder diffractometer DMC.
	(bottom) data from the triple-axis spectrometer \mbox{RITA-2}, 
	using energy analysis of the scattered neutrons.
        \label{fig:Tdiffraction}}
\end{figure}

Fig.~\ref{fig:Tdiffraction} shows the results of the diffraction experiments. The AFM hematite peaks (003) and (101) are clearly
visible at $q$-values of 1.37~\AA$^{-1}$ and 1.51~\AA$^{-1}$, respectively, in both data sets. 
However, the temperature dependence of the magnetic scattering intensity is most different in the two experiments. 
In the two-axis experiment the  peak intensity at 300~K is reduced by 10\% relative to the value at 10~K, whereas
the triple-axis experiment shows a reduction of 90\%. The only difference between these two experiments is the energy resolution,
indicating that the hematite spins become almost completely dynamic with an energy scale in the range between
the respective energy resolutions of the two instruments.

\begin{figure}[]
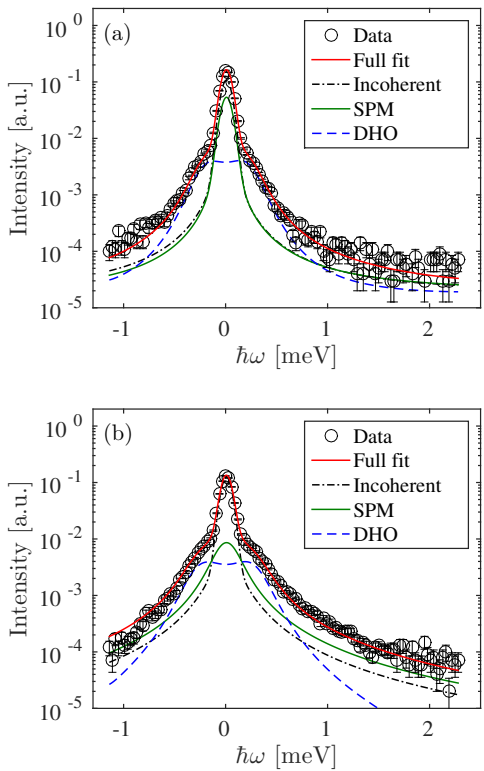

{        \includegraphics[width=0.35\textwidth]{RITA2_mag_50K.eps}} \\ \vspace{5mm} 
{        \includegraphics[width=0.35\textwidth]{RITA2_mag_150K.eps}}
\caption{
        (color online) Inelastic neutron scattering data at
        $q=1.37$~\AA$^{-1}$, corresponding to the (003) AFM Bragg peak.
	measured at 50~K (top) and 150~K (bottom).
        Counts are normalized to monitor and plotted on logarithmic axis
        vs.\ energy transfer, $\hbar\omega$.
        The solid red lines show the best fit to the model (\ref{eq:DHO})
        convoluted with experimental broadening.
	Individual components of the fits are shown by 
	dot-dashed black lines, solid green lines, and dashed blue lines, which 
	represent the incoherent elastic background, the quasielastic SPM signal, 
	and the DHO component, respectively.
        \label{fig:Tscans}}
\end{figure}

These initial observations prompted us to re-analyse previously published inelastic neutron scattering data from the same sample 
in the temperature range 10-200~K \cite{kuhn06}, as well as unpublished data up to 300~K.
Fig.~\ref{fig:Tscans} shows our data, 
displaying the signal at a scattering vector of $q = 1.37$~\AA$^{-1}$, 
corresponding to the AFM (003) reflection.
The strongest feature is the central elastic peak that stems from elastic incoherent
and quasielastic background and a narrow quasielastic SPM signal. 

The uniform modes are seen in the data as broad side peaks at $|\hbar \omega_\alpha| \sim 0.3$~meV.
We model these peaks by a damped harmonic oscillator (DHO) function \cite{hansen97,klausen04}.
Before correcting for experimental resolution and constant background,
the model reads
\begin{eqnarray} \label{eq:DHO} \lefteqn{
        I(\omega)=A_{\rm inc}\delta(\omega)+C} && \\ &+&
        \frac{A_{\rm qe}  \Gamma_{\rm qe}/\pi}{\Gamma_{\rm qe}^2 + \omega^2} +
        \frac{A_{\rm SPM}  \Gamma/\pi}{\Gamma^2 + \omega^2} +
        \frac{ 2 A_{\rm DHO} D(\omega)  \omega_\alpha^{2} \gamma_\alpha/\pi}
        {(\omega^2-\omega_\alpha^2)^2+4\gamma_\alpha^{2}\omega^2}, \nonumber
\end{eqnarray}

where $\hbar \omega$ is the neutron energy transfer, 
$A_{\rm inc}$, $A_{\rm qe}$, $A_{\rm SPM}$, and $A_{\rm DHO}$ are the integrated intensities of the
incoherent elastic, the quesielastic incoherent (from mobile H$_2$O), the quasielastic magnetic (from SPM),
and the magnetic DHO components, respectively.
$\Gamma_{\rm qe}$, $\Gamma$, and $\gamma_\alpha$ are the widths (HWHM)
of the quasielastic incoherent, the quasielastic magnetic, and the inelastic peaks, respectively.
$D(\omega)=\omega [n(\omega)+1]/[k_{\rm B}T]$
is the detailed balance factor \cite{hansen97}, 
with $n(\omega)$ being the Bose factor and $k_{\rm B}$ being the Boltzmann constant,
as detailed in \cite{kuhn06}.
The values of $A_{\rm inc}$, $A_{\rm qe}$, and $\Gamma_{\rm qe}$ are found from 
constant-$q$ scans at non-magnetic $q$-values, 
leaving seven free fitting parameters for the magnetic scattering:
$A_{\rm SPM}$, $\Gamma$, $A_{\rm DHO}$, $\omega_\alpha$, $\gamma_\alpha$, $C$ and a possible offset of the elastic line due to imprecision in alignment.

\begin{figure}[]
\includegraphics[width=0.35\textwidth]{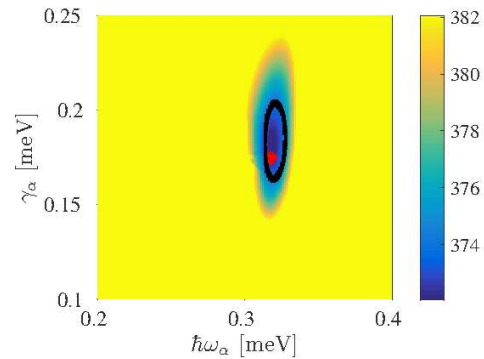}
\caption{ (color online)
       The quality of the model fit to the inelastic data, measured as the unreduced chi-square value, 
       $\chi_{\rm unred}^2$, shown for the parameter
       plane spanned by $\omega_\alpha$ and $\gamma_\alpha$. At each point in this plane, the value of the five other free fitting parameters have been fitted to the 115 data points. The 95\% confidence interval, corresponding to 
       $\Delta \chi_{\rm unred}^2 = 2.3$, is marked by the solid black line.
\label{fig:chisquare}}
\end{figure}

\begin{figure}[]
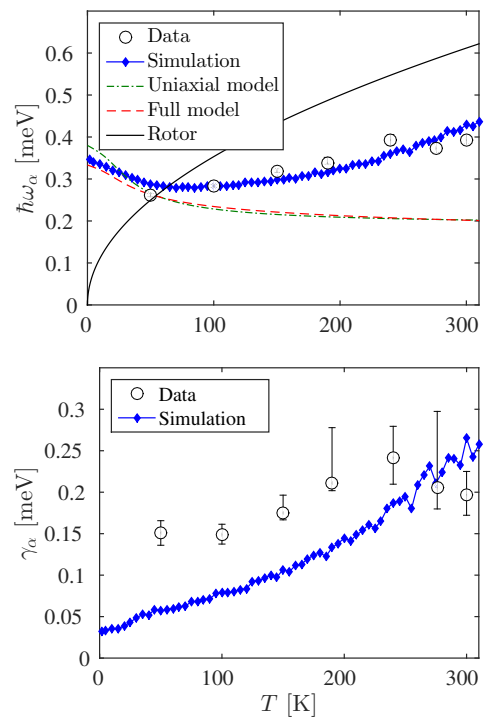

\includegraphics[width=0.35\textwidth]{omega_t_dep.eps}\\
\includegraphics[width=0.35\textwidth]{gamma_dho_t_dep.eps} 
\caption{ (color online)
        Position ($\omega_\alpha$) (top), and width ($\gamma_\alpha$) (bottom)
        of the inelastic DHO signal as a function of
        temperature.
Black circles are measured values and blue diamonds are results from the Langevin simulations, described in the text.
        Error bars on the simulated data points are comparable to the symbol sizes.
	In the top panel, the dot-dashed green curve in the $\hbar \omega_\alpha$ plot is the previously published result for the uniform mode \protect\cite{hansen97}, the dashed red line is the 
	more accurate version of
	the same calculation \protect\cite{Hansen2000},
	and the solid black curve is our analytical result for the ''rotor`` mode, 
	eq.\ (\protect\ref{eq:wrot}).
\label{fig:Tparams}}
\end{figure}

Experimental data taken at 10~K have intrinsically low signal-to-noise
ratio. The parameter values for the inelastic signal then becomes unreliable, since the fit tends to model 
non-Gaussian tails in the resolution function for the large elastic signal, rather than the tiny inelastic magnetic signal. It was therefore necessary to disregard data taken at this temperature in the data analysis.

The quality of the fit of the model (\ref{eq:DHO}) to data is crucial to this work. 
Some correlation could be expected between the free parameters. For the important DHO signal,
two parameters are of the same energy scale of a few tenths of a meV: $\gamma_\alpha$ and $\omega_\alpha$,
and we have therefore investigated how these two parameters correlate by fixing their values and performing
a fit of the other five parameters. Then, we record the goodness of fit given by the value of the unreduced 
chi-square function, $\chi_{\rm unred}^2$ function.
We have done this for a range of parameters, spanning the $(\gamma_\alpha, \omega_\alpha)$-plane with the results
presented in Fig.~\ref{fig:chisquare}. The confidence interval in such a figure is given by
the condition\cite{recipes} $\Delta \chi_{\rm unred}^2 < 2.3$  and is marked by the black solid line in the figure. 
We see that the values of the two parameters are essentially uncorrelated and that the fitted parameters 
are robust; except that the $\gamma_\alpha$ parameter may have a somewhat asymmetric errorbar. 

Our central findings of experiments and model fit, the energy of 
the uniform mode $\hbar\omega_\alpha$, is shown in fig.~\ref{fig:Tparams}.
$\omega_\alpha$ is found to increase with temperature, 
in direct contrast to the results from 15~nm particles \cite{hansen97}.

\begin{figure}[]
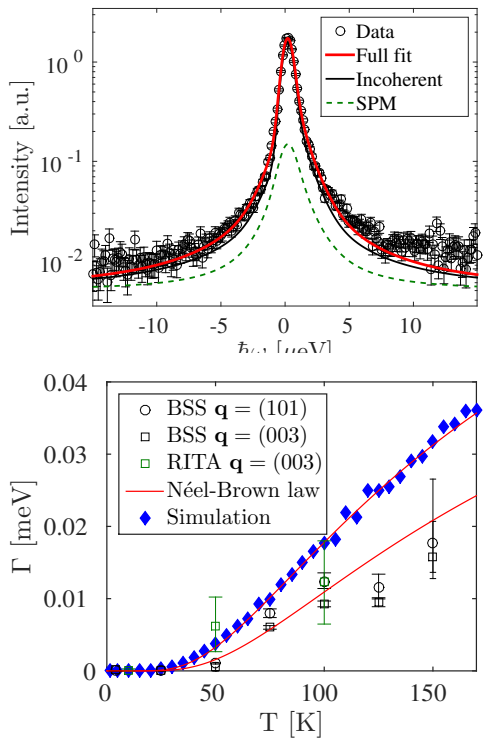

         \includegraphics[width=0.35\textwidth]{Coated_detno5_50K.eps}
         \includegraphics[width=0.35\textwidth]{8nm_widths.eps}
 \caption{
        (Color online) quasielastic neutron scattering data from BSS, J\"ulich, 
        taken close to the magnetic (003) AFM Bragg peak.
        (top) raw data taken at $T=50$~K with a Lorentzian model fit to data as explained in the text. 
        (bottom) Fitted Lorentzian line widths of the superparamagnetic signal as a function of temperature.
        The solid lines are fits to the Ne\`el-Brown law as described in the text.
        \label{fig:QENS}}
\end{figure}

The width of the central peak was further studied by high-resolution quesielastic measurements around 
the magnetic (003) peak, addressing superparamagnetism fluctuations. One example of
the raw data is presented in fig.~\ref{fig:QENS} (top) together with a model fit that contains a Lorentzian line shape convoluted with the resolution function, plus a constant background. 
In fig.~\ref{fig:QENS} (bottom), the Lorentzian linewidths are plotted as a function of temperature. 
The width is seen to increase with temperature up to around 150~K, 
where the magnetic signal becomes wider than the instrument energy window and no useful information can be extracted. 
The data up to 150~K is well modeled by the Ne\`el-Brown law (\ref{eq:neelbrown}),
giving an anisotropy value of $KV/k_{\rm B} =194(40)$~K with $\tau_0=9(2)\times 10^{-12}$~s. 
   
The remainder of this article
is devoted to the explanation and modeling of the two unforeseen results:
The vanishing of the diffraction peak at room temperature in the triple-axis experiments
and the increase of precession frequency with temperature.
In the discussions, we will also include the quasielastic neutron scattering data. 

\section{Analytical theory}
A derivation for the spectrum of a fluctuating ferromagnetic nanoparticle in a uniaxial anisotropy was earlier
derived by W\"urger \cite{wurger98}. However, in this scenario, the frequency of the uniform modes decrease with increasing temperature, in contrary to our observations. For this reason, we decided to develop an analytical understanding for this AFM system with two anisotropies.

The generally accepted microscopic magnetic Hamiltonian for hematite nanoparticles reads \cite{bahl08}:
\begin{equation} 
\mathcal{H} = \sum_{ij} J_{ij} {\bf s}_i \cdot {\bf s}_j
    - \kappa_1 \sum_i (s_i^x)^2
    - \kappa_2 \sum_i (s_i^z)^2, \label{eq:hamilton} 
\end{equation}
where $i$ and $j$ are atomic indices, 
$\kappa_1 < 0$ is a large planar anisotropy, and
$\kappa_2 > 0$ is a smaller axial anisotropy, which is linked to the anisotropy barrier in Eq.~\eqref{eq:neelbrown} through 
$KV/k_{\rm B} = \kappa_2ss'N$, where $s' = s-1/2$ \cite{jacobsenthesis}.
$J_{ij}$ is the exchange coupling constants.
In this notation, $y$ and $z$ lie within the basal hematite $(a,b)$ plane, while $x$ is along
the $c$-axis.

To describe the low-energy dynamics,
we follow W\"urger \cite{wurger98} and assume collective motion ($q=0$) of the spins in each of the
two sublattices (${\rm A}$~and~${\rm B}$). This is justified by the fact that the highest temperature in our study, 300~K, is far below the N\'eel temperature of the system and therefore the spin waves have reduced the ordered moment of the sublattices only slightly. In addition, the finite size of the system results in a quantization of the spin wave spectrum, and due to the steepness of the hematite spin wave dispersion, only few spin waves states are in fact allowed below energies corresponding to 300~K \cite{Hansen2000}. 

We define the number of spins in the two sublattices as, $N_A = N_B $ and define the uncompensated fraction $\xi = N_A / N_B$.
We initially consider the case $\xi=1$.
The spin dynamics can be fully described by the two superspins
\begin{equation} \label{eq:superspin}
{\bf S}_{\rm A} = \hbar\sum_{i \in \rm A} {\bf s}_i \quad , \quad
{\bf S}_{\rm B} = \hbar\sum_{i \in \rm B} {\bf s}_i .
\end{equation}
In terms of ${\bf S}_{\rm A}$ and ${\bf S}_{\rm B}$, the Hamiltonian becomes
\begin{eqnarray}
\mathcal{H} &=& J' {\bf S}_{\rm A}\cdot {\bf S}_{\rm B} - \kappa_1' ({S_A^x}^2+{S_B^x}^2) \nonumber \\
&& - \kappa_2'({S_A^z}^2+{S_B^z}^2)
-\gamma {\bf B} \cdot({\bf S}_{\rm A} + {\bf S}_{\rm B}),
\end{eqnarray}
To simplify, we define $\hat{\bf S}_{\rm A}={\bf S}_{\rm A}/S$, and similarly for $\hat{\bf S}_{\rm B}$. 
We then reach
\begin{eqnarray} \label{eq:hamiltonB}
 \mathcal{H} &=& \gamma S [ B_\mathrm{ex} \hat{\bf S}_{\rm A} \cdot \hat{\bf S}_{\rm B}   +\frac{B_1}{2} ({\hat{S}_{\rm A}^x}{}^2+{\hat{S}_{\rm B}^x}{}^2  ) \nonumber \\
&&-\frac{B_2}{2} ({\hat{S}_{\rm A}^z}{}^2+{\hat{S}_{\rm B}^z}{}^2)], 
\end{eqnarray}
where the gyromagnetic ratio is $\gamma = g \mu_{\rm B}/\hbar$ and 
$B_{\rm ex} \gg B_1 > B_2$.
Here, $B_{1}$ and $B_{2}$ are the planar and uniaxial anisotropy fields, respectively, defined by 
$g \mu_{\rm B} B_i = 2|\kappa_i| s'$ for $i=1,2$. The exchange field is 
$g \mu_{\rm B} B_{\rm ex} = 2zJ_{\rm nn}s$, where the factor of 2 is due to double-counting.

In bulk hematite,  $B_\mathrm{ex}=827$~T \cite{samuelsen1970,klausen04,morrish}. From the neutron scattering data presented in Fig.~\ref{fig:Tscans} and the model Eq.~\ref{eq:uniform1} we find $B_2=4.8$ mT. $B_1$ cannot be determined from the present neutron scattering data and we therefore use the value $B_1(\text{8 nm}) = B_1(\text{16 nm})$ \cite{hill14}, $B_1 \approx 46$~mT at low temperatures, gradually increasing to $\approx 66$~mT at 300~K.





The exchange term is minimal
when $\hat{\bf S}_{\rm A}$ and $\hat{\bf S}_{\rm B}$ point in opposite directions;
the $B_1$ anisotropy is minimized if they lie in the $(y,z)$ plane, and the $B_2$ anisotropy 
will attain its minimum when $\hat{\bf S}_{\rm A}$ and $\hat{\bf S}_{\rm B}$ are parallel to the $z$-axis. 
Hence we have two minimum energy configurations $\hat{\bf S}_{\rm A} = (0,0,\pm 1)$, $\hat{\bf S}_{\rm B} = (0,0,\mp 1)$. 


Since the magnitudes of the superspins are much larger than $\hbar$, 
quantum mechanics plays no role, and the dynamics is governed by 
the classical equations of motion: 
\begin{equation}
\frac{\mathrm{d}S^i}{\mathrm{d}t}= [S^i,\mathcal{H}],
\end{equation}
where the right hand side is a Poisson bracket, obeying the rule
$[S^x, S^y] = S^z$, etc. Using $\mathcal{H}$ from (\ref{eq:hamiltonB}), we get

\begin{equation}
\frac{\mathrm{d}\hat{\bf S}_{\rm A}}{\mathrm{d}t} = \gamma \hat{\bf S}_{\rm A} \times {\bf B}_{\rm A} , \; \frac{d\hat{\bf S}_{\rm B}}{dt} = \gamma \hat{\bf S}_{\rm B} \times {\bf B}_{\rm B} ,
\end{equation}
where ${\bf B}_{\rm A}$ is the resulting field on sublattice A originating partly from the
anisotropies, partly from the magnetisation of sublattice B:

\begin{eqnarray} 
\mathbf{B}_{\rm{A}} &=& -~  B_{\rm ex} \hat{\mathbf{S}}_{\rm B} 
  +~  B_{1} \hat{S}_{\rm A}^x \hat{\bf e}^x+
B_{2} \hat{S}_{\rm A}^z \hat{\bf e}^z,  \nonumber \\
\mathbf{B}_{\rm{B}} &=& -~  B_{\rm ex} \hat{\mathbf{S}}_{\rm A} 
  +~  B_{1} \hat{S}_{\rm B}^x \hat{\bf e}^x+
B_{2} \hat{S}_{\rm B}^z \hat{\bf e}^z, \label{eq:B_eff} 
\end{eqnarray}
 The scalar $\hat{S}_{A}^x$ is the projection of $\hat{\bf S}_{\rm A}$ 
onto the $x$-axis.
Considering only motion with small deviations from the minimum energy configuration, 
we obtain two normal modes with frequencies \cite{bahl08}

\begin{align}
\omega_\alpha &= \gamma \sqrt{ (2B_\mathrm{ex}+B_1+B_2)    B_2    }\approx \gamma \sqrt{ 2B_\mathrm{ex}    B_2    } \label{eq:uniform1} \\ 
\omega_\beta &= \gamma \sqrt{ (2B_\mathrm{ex} +B_2)(B_1+B_2)       } \approx\gamma \sqrt{    2B_\mathrm{ex} (B_1+B_2)     }. \nonumber
\end{align}
These are the two uniform modes calculated earlier, and also observed directly with inelastic
neutron scattering \cite{hansen97,klausen04, hill14}. The model can be extended to include $\xi \neq 1$, 
see Ref.~\onlinecite{bahl08}.
%
%

The temperature dependence of the excitation energies at low temperatures is given by
\begin{align}
 \omega_{\alpha,\beta}(T)=\omega_{\alpha,\beta}(0) \langle \hat{S}^z \rangle, \label{eq:T_dep}
\end{align}
where the ordered sublattice moment $\langle \hat{S}^z\rangle$ decreases monotonically with temperature,
according to Boltzmann statistics \cite{klausen04,hill14}.

\begin{figure}[]
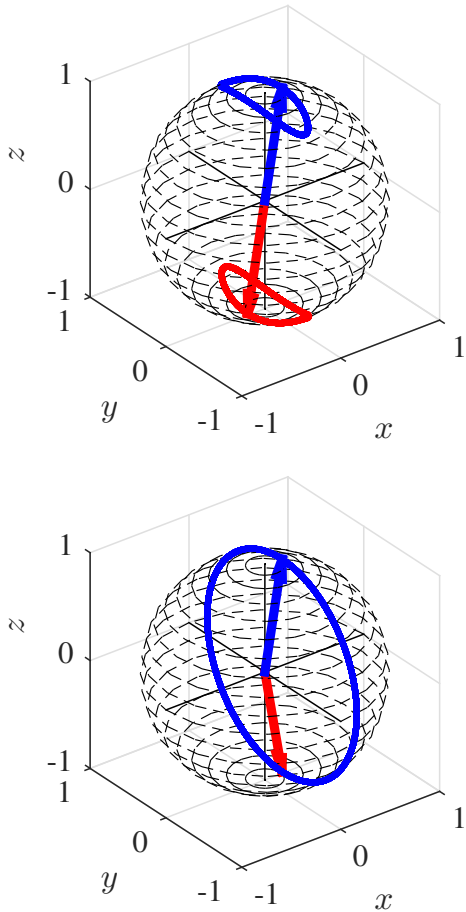

  \includegraphics[width=0.40\textwidth]{CME.eps}
  \includegraphics[width=0.40\textwidth]{RM.eps}
  \caption{(Color online) Illustration of two magnetic modes in an antiferromagnetic nanoparticle with a vertical easy axis. The A (B) sublattice points in the $+z$ ($-z$) direction and describes
  a blue (red) trajectory:
a) a uniform in-plan magnetic mode; the $\omega\alpha mode$;
b) the new ''rotor mode``. }
        \label{fig:modes}
\end{figure}

It turns out that the system supports another approximate mode, which we will denote the ``rotor'' mode,
described in the following. 
We assume that the magnetic moments move slightly out of the $(y,z)$ plane, 
but not with strictly antiparallel sublattices, {\em e.g.}
\begin{equation}
\hat{\bf S}_{\rm A} = ( \theta_{\rm A},\hat{S}_y, \hat{S}_z)\quad {\bf S}_{\rm B} = (\theta_{\rm B}, -\hat{S}_y,-\hat{S}_z).
\end{equation}
We first note that the canting angles described here are out of the basal plane, 
and therefore unrelated to the in-plane canting angle caused by the DM-interaction. 
With our {\em ansatz}, the equations of motion become

\begin{eqnarray}
\frac{\mathrm{d}\theta_{\rm A}}{\mathrm{d}t} &=&  \gamma B_2  \hat{S}_y \hat{S}_z \nonumber \\ \frac{\mathrm{d}\theta_{\rm B}}{\mathrm{d}t} &=& \gamma B_2  \hat{S}_y \hat{S}_z  \nonumber \\
\frac{\mathrm{d}\hat{S}_y}{\mathrm{d}t} &=& - \gamma B_\mathrm{ex} (\theta_{\rm A}+\theta_{\rm B}) \hat{S}_z \nonumber \\
\frac{\mathrm{d}\hat{S}_z}{\mathrm{d}t} &=& \gamma B_\mathrm{ex}(\theta_{\rm A}+\theta_{\rm B}) \hat{S}_y, 
\end{eqnarray}
where we neglect terms of order $B_1$ in comparison to terms of order $B_\mathrm{ex}$.
Assuming that the angles $\theta_{\rm A}$ and $\theta_{\rm B}$ are constant in time, 
the magnetic moments will perform full $2\pi$ rotations around the $x$-axis with a frequency given by
\begin{equation} \label{eq:wrot}
\omega_{\rm rot} = \gamma B_\mathrm{ex}(\theta_{\rm A}+\theta_{\rm B}).
\end{equation}
This rotation is illustrated in Fig.~\ref{fig:modes} for the case $\theta_{\rm A}=\theta_{\rm B}$. We denote the sublattice canting angle $\theta = |\theta_{\rm A}+\theta_{\rm B}|$. 
The $\theta$ dependence of the overall system energy stems from the exchange interaction,
$E_{\theta} \propto J S^2 \cos(\theta)$.
From Boltzmann statistics, we can now calculate the thermal average value of $\theta^2$ to
\begin{equation}
\langle \theta^2\rangle\approx2k_B T/(\gamma SB_\mathrm{ex}) .
\end{equation}

We next investigate the assumption
that the relative rate of change of the angles, $\dot{\theta}/\theta$, 
is small compared to $\omega_{\rm rot}$.  Neglecting thermal fluctuations we obtain
\begin{equation} \label{eq:orderofmagn}
\frac{\dot{\theta}/\theta}{\omega_{\rm rot}} \approx \frac{B_2}{2B_\mathrm{ex} \theta^2} \approx 5\times 10^{-3}\frac{N_A}{T [K]},
\end{equation}
where we have used the values of $B_2$ and $B_{\rm ex}$ given in the following.
For nanoparticles corresponding to 8~nm hematite, $N_A \sim 5000$, and for $T \sim 100$~K, (\ref{eq:orderofmagn}) 
gives $\dot{\theta}/(\theta \omega_{\rm rot}) = 0.3 < 1$.
Hence, $\theta_{\rm A}$ and $\theta_{\rm B}$ maintain their values for sufficiently long times
that the rotor mode is significant.

\section{Numerical simulations and results}
To veryfy the soundness of the analytical approximations, and to obtain additional detail in the model,
we numerically simulate the sublattice dynamics, using Langevin dynamics.
Here, the equation of motion is a stochastic Landau-Lifschitz equation,
very similar to the method applied earlier to describe superparamagnetism 
in ferromagnetic nanoparticles \cite{langevin}. In the numerical calculations, we
have lifted the requirement $\xi=1$ and also included the  Dzyaloshinsky--Moriya interaction.
In this scheme, the equations of motion for the two sublattices become:
\begin{eqnarray} \label{eq:langevin}
\frac{\mathrm{d}}{\mathrm{d}t} \hat{\bf S}_{\rm A}
  &=& \gamma \hat{\bf S}_{\rm A} \times (\hat{\bf B}_{\rm A} + {\bf b})
    - \lambda|\gamma| \hat{\bf S}_{\rm A} \times
      (\hat{\bf S}_{\rm A} \times {\bf B}_{\rm A})  \\
\frac{\mathrm{d}}{\mathrm{d}t} \hat{\bf S}_{\rm B}
  &=& \gamma \hat{\bf S}_{\rm B} \times (\hat{\bf B}_{\rm B} + {\bf b})
    - \lambda|\gamma| \hat{\bf S}_{\rm B} \times
      (\hat{\bf S}_{\rm B} \times {\bf B}_{\rm B}),\nonumber
\end{eqnarray}
where
%
\begin{eqnarray} 
\mathbf{B}_{\rm{A}} &=& -~  \xi B_{\rm ex} \hat{\mathbf{S}}_{\rm B} 
  +~  B_{1} \hat{S}_{\rm A}^x \hat{\bf e}^x+ \nonumber \\
 && B_{2} \hat{S}_{\rm A}^z \hat{\bf e}^z-\xi B_D(\hat{S}_B^z \hat{\bf e}^y-\hat{S}_B^y\hat{\bf e}^z),  \nonumber \\
\mathbf{B}_{\rm{B}} &=& -~  B_{\rm ex} \hat{\mathbf{S}}_{\rm A} 
  +~  B_{1} \hat{S}_{\rm B}^x \hat{\bf e}^x+ \nonumber \\
  && B_{2} \hat{S}_{\rm B}^z \hat{\bf e}^z+ B_D(\hat{S}_A^z \hat{\bf e}^y-\hat{S}_A^y\hat{\bf e}^z) 
\end{eqnarray}
$B_D$ is the Dzyaloshinsky--Moriya field, where we use the bulk hematite value of $B_D=2.1$~T\cite{morrish}.
In (\ref{eq:langevin}), {\bf b} is a small random field representing thermal fluctuations,
which are isotropic and uncorrelated in time. The term $\lambda$ represents dissipation.
The fluctuation-dissipation theorem \cite{fluct_diss} is used to determine 
in absolute units the simulated temperature:
\begin{equation} \label{eq:fluct_diss}
k_{\rm B} T =   |\gamma| g\mu_B s \frac{\langle {\bf b}_A^2 \rangle N_A}{2\lambda}.
\end{equation}
%

Note that the particle size is present here through $N_A$. The fluctuating field is uncorrelated in time, $\langle b(t) b(t')\rangle\propto \delta(t-t')$.

The equations of motion (\ref{eq:langevin}) were integrated numerically
using the Runge-Kutta method of second order \cite{recipes} with a time step of
1~fs (1000~THz), much faster than typical frequencies of the uniform mode $\omega_\alpha$
(0.5 THz). 
The method was tested against analytical results
in the limit of vanishing fluctuation and dissipation \cite{bahl08,morrish}, and for the
one-sublattice (ferromagnetic) case \cite{garde08,jacobsenthesis}.

\begin{figure}[]
        \includegraphics[trim = 0mm 0mm 0mm 0mm, clip, width=0.45\textwidth]{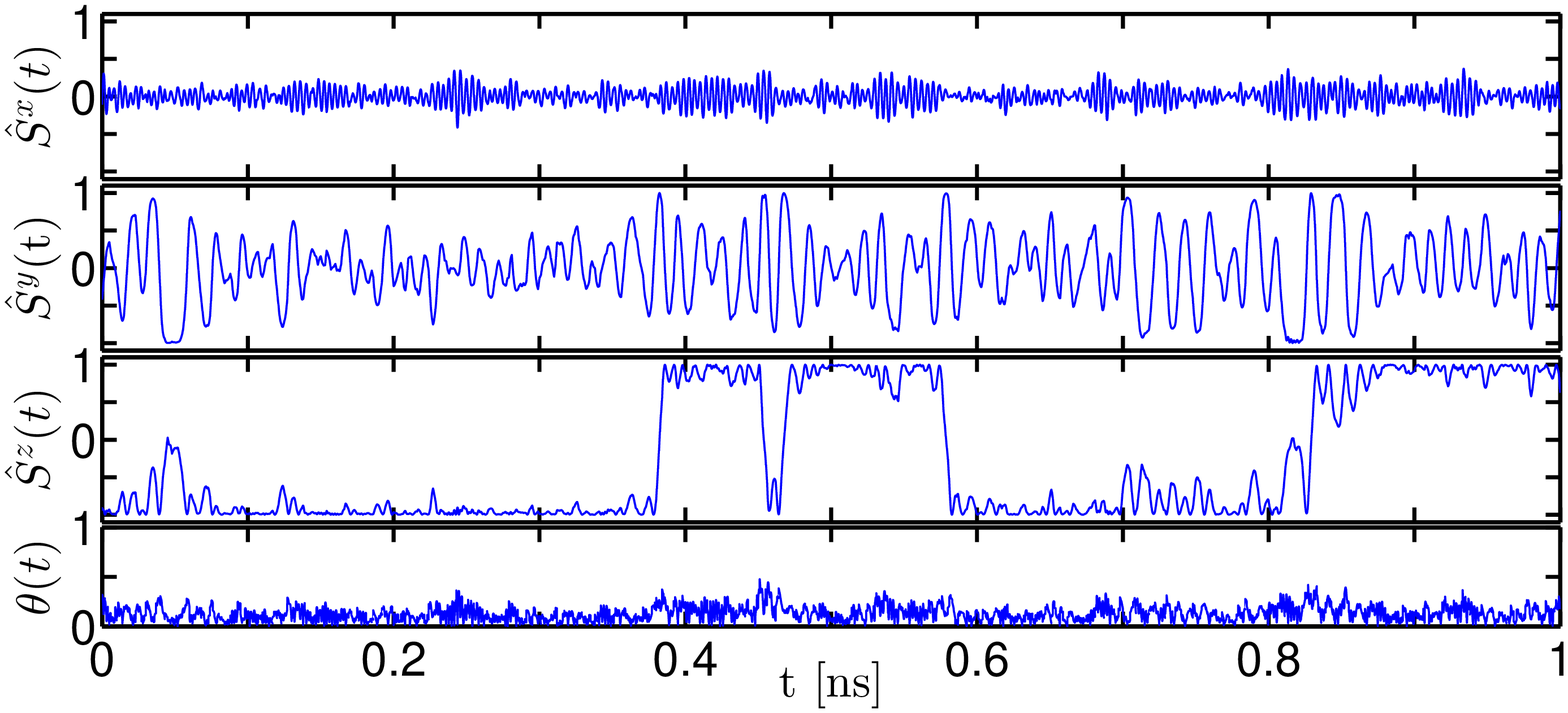}
        \includegraphics[trim = 0mm 0mm 0mm 0mm, clip, width=0.45\textwidth]{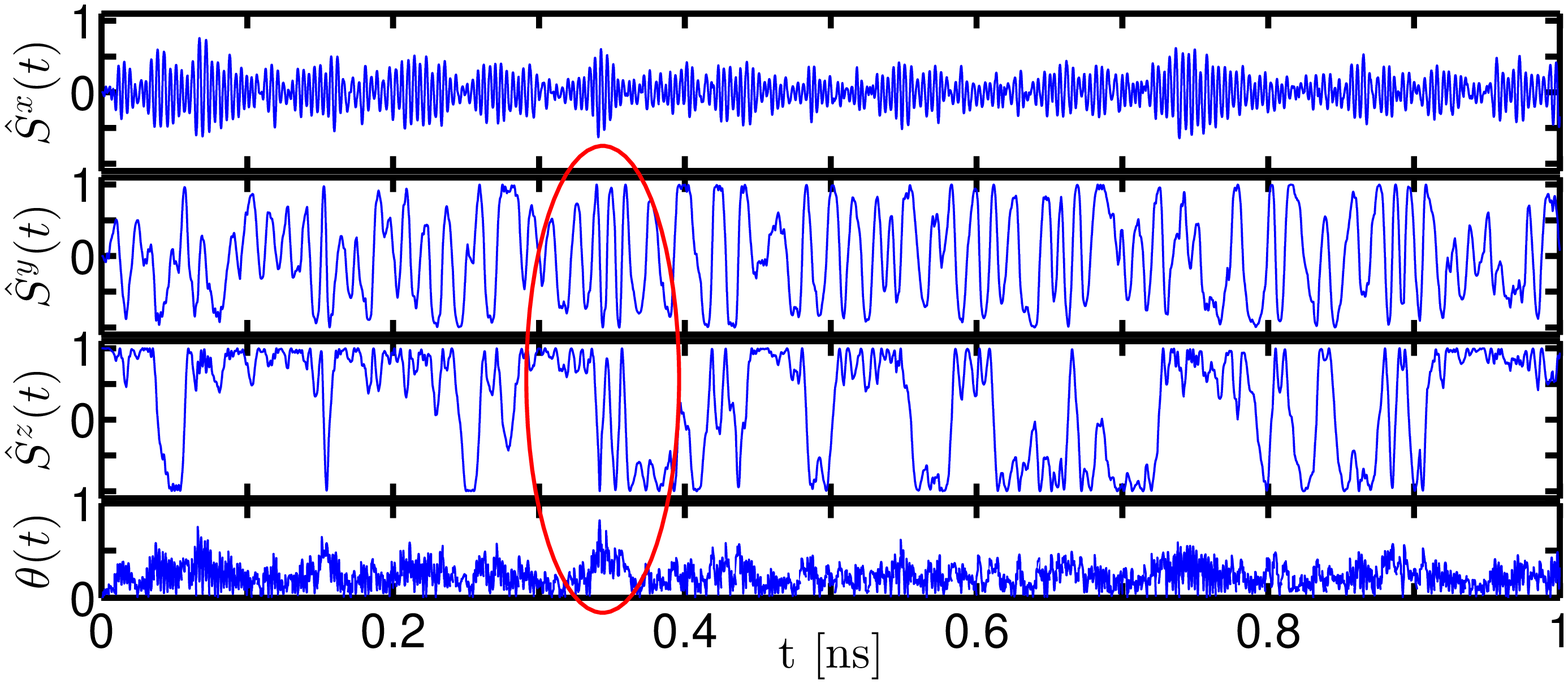}
        \includegraphics[trim = 0mm 0mm 0mm 0mm, clip, width=0.45\textwidth]{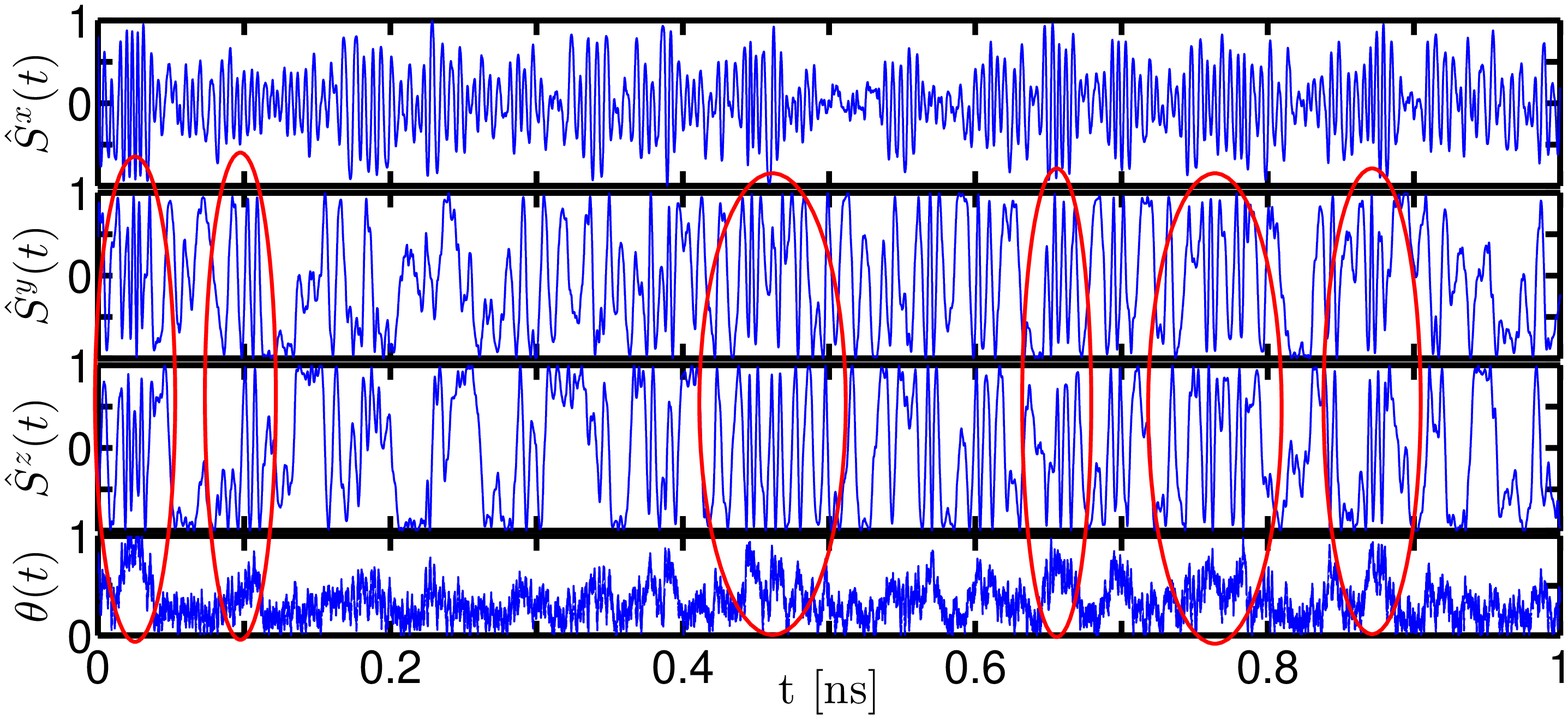}
\caption{(color online) Raw data from the Langevin simulations, showing the time development
  of the three spin components at temperatures of 40~K (top), 150~K (center) and 300~K (bottom).
  $\theta(t)$ represents the canting and is displayed in degrees.
  The encircled areas in the center and bottom panel show fast coherent oscillations in the $yz$-plane 
  correlated with a large value of the canting angle.
\label{fig:rawsim}}
\end{figure}

\begin{figure}[]
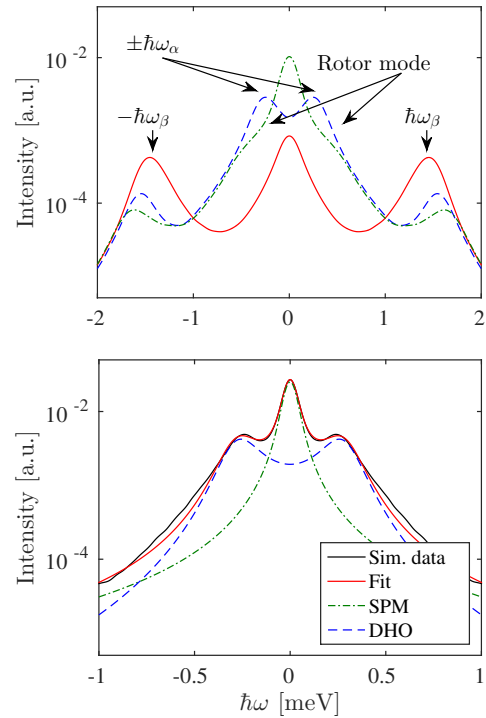

\includegraphics[width=0.35\textwidth]{sim_xyz_example.eps}
\includegraphics[width=0.35\textwidth]{sim_xyz_example_fit.eps}
\caption{(color online) Power spectra from the Langevin simulations at 150~K.
  Top: The individual components, $\hat{S}_x$ (solid red), $\hat{S}_y$ (dashed blue), and $\hat{S}_z$ (dash-dotted green). All three oscillator modes are clearly visible, as indicated in the figure. 
  Bottom: Sum of the $y$ and $z$ components (solid black) with a fit (solid red). 
  The individual components of the fit are the SPM and the DHO signals, given by
  the dash-dotted green and dashed black lines, respectively.
\label{fig:powersim}}
\end{figure}

Examples of raw simulation data
for the two-sublattice case are shown in Fig.~\ref{fig:rawsim}.
The motions of the sublattice magnetizations were, in turn, Fourier transformed to obtain
the power spectra, which through the Wiener-Khintzine formula \cite{WK} 
yield the simulated scattering intensities.
Since the scattering vector lies in the $x$-direction,
we sum the power spectra of the $y$ and $z$ spin components to model experimental data.
The simulated data was convoluted with the experimental resolution and  fitted in the same way
as the experimental data, using (\ref{eq:DHO}), 
resulting in good agreements over three orders of magnitude; 
see Fig.~\ref{fig:powersim}. 
The deviations found at the small-amplitude data at $|\hbar \omega| >  0.8$~meV 
are irrelevant for our conclusions and we neglect them in the following.

In the simulations we have adjusted four parameters: The uniaxial anisotropy field $B_2$, the exchange field $B_{\rm ex}$, the uncompensated moment fraction, $\xi$ and the dissipation parameter $\lambda$ to find the values that yield
the best agreement between simulations and experiment with respect to the temperatures dependences 
of $\gamma_\alpha$ and $\omega_\alpha$ \cite{garde08}. 
We find $B_2=9.5$~mT, corresponding to $KV/k_{\rm B} = 169$~K. 
$B_X$ and $\xi$ cannot be determined independently: any value of $B_X$ between 600 and 1000~T has a corresponding value of $\xi$ where simulations and data match. We therefore fix the exchange field to the bulk value, $B_X=827$~T, and thus find $\xi=1.010$. Finally, we determine $\lambda=6\times 10^{-4}$.

\section{Discussion}

\subsection{Existence of the rotor mode}
We first discuss our hypothesis of the existence of a rotor mode by comparing data from
neutron scattering experiments, numerical simulations, and analytical calculations.

Fig.~\ref{fig:Tparams} (top) displays the temperature dependence of
the mean precession frequency, $\omega_\alpha$, defined through (\ref{eq:DHO}).
The agreement on the temperature behaviour of $\omega_\alpha$ 
between experiment and simulation is excellent.

The value of the DHO peak width $\gamma_\alpha$, displayed in Fig.~\ref{fig:Tparams} (bottom), is generally higher 
in the experimental data than in the simulations. 
We ascribe the higher experimental value of $\gamma_\alpha$
to a distribution of particle sizes and shapes,
which results in a spread in $\omega_\alpha$. 
This would appear in the fit as an overall increase of $\gamma_\alpha$, most prominent at the lowest
values of this parameter, as the particle size distribution effect is presumably constant and the width
of the two broadenings should add in quadrature.
This is in good overall agreement with the experimental observations.

The simulated values of $\omega_\alpha$ are in excellent agreement with the neutron scattering data. At low temperatures, $T \leq 50$~K, the analytical uniform mode prediction agrees well with the simulations. At higher temperatures, the uniform mode prediction strongly underestimates the value of $\omega_\alpha$, while our analytical results for the rotor mode overestimates this value. We believe the reason for this to be the following: Both the uniform mode and the rotor mode give rise to a DHO signal, which due to the relatively short life time of the excitations are both quite broad.  Experimentally we cannot resolve the two individual DHOs, and a single DHO has therefore been used to fit the data. The observed frequency is thus a weighted average of the frequency of the rotor mode, $\omega_{\rm rot}$ and that of the uniform mode, $\omega_\alpha$, in general agreement with the experimental and simulated data. 


Additional insight is found by inspecting the simulated time evolution of the magnetic dynamics,
shown in Fig.~\ref{fig:rawsim}. The 40~K data show
low-amplitude, high-frequency oscillations in $\hat{S}^x$ and high-amplitude,
low-frequency oscillations in $\hat{S}^y$, representing the two uniform modes 
(\ref{eq:uniform1}).
The easy-axis component, $\hat{S}^z$, displays nearly constant values,
interrupted by sudden (SPM) magnetization reversals.
These three observations are in accordance with the present understanding of magnetic dynamics
of nanoparticles.
The 150~K data contains periods with much less regular motion of the sublattice spins. In particular,
both $\hat{S}^z$ and $\hat{S}^y$ show oscillations.
The highlighted regions show time intervals with coupled fast oscillations
in these parameters, correlated with a high value of $\theta$. These fast oscillations can be seen as repeated, coherent SPM relaxations and rarely appear at the lowest temperatures. 
At 150~K they appear at least once per simulated nanosecond,  and  dominate at larger temperatures, {\em e.g.} at 300~K.
These events are indeed occurences of short-lived rotor dynamics.

Our explanations are supported by the power spectra shown in Fig.~\ref{fig:powersim}, where 
the SPM behaviour is found in spin components along the easy $z$-direction, 
the DHO signals are seen in the $y$
and $x$ components, while the rotor mode is seen as a high-frequency shoulder in the $z$ signal.
The expected rotor contribution to the spectrum of $S_x$ is not clearly visible due to the broadening
of the signal from the $\omega_\alpha$ mode.


\subsection{Comparison with earlier work}
In earlier work by our group, we found a value for the energy barrier $KV/k_{\rm B} = 250(30)$~K using the ratio between elastic and inelastic signal in the neutron scattering data from the same sample, and $KV/k_{\rm B} = 335(35)$~K using M\"ossbauer spectroscopy data of the same sample \cite{kuhn06}. However, this work used an effective 3D model, in which the in-plane anisotropy is neglected, leading to an overestimation of $KV$. Assuming that the in-plane anisotropy is identical to the value for 16~nm particles \cite{hill14}, and that $\xi=1, $the values become 
$KV/k_{\rm B} = 148(18)$~K from the neutron data and $KV/k_{\rm B} = 211(20)$~K from the M\"ossbauer data. From the values of $\omega_\alpha$, we furthermore found an uncompensated moment of $\xi=1.010$ using the uniaxial model.

From the fit of the backscattering data to the Ne\`el-Brown model, 
we find in the present work a value of the energy barrier, $KV$ of 194(40)~K, in agreement with the
M\"ossbauer data. In addition we obtain the prefactor value $\tau_0^\text{sim}=9(2)\times10^{-12}$~s.

Tuning the simulations to obtain best agreement between the measured and simulated values of $\hbar \omega_\alpha$ and $\gamma_\alpha$, we reach a "modeled refinement" of the neutron data, gving $KV/k_{\rm B} = 169$~K and $\xi=1.010$, in very good agreement with earlier work. The simulated backscattering data using the same simulation parameters also follow the N\'eel-Brown law very closely, leading to $KV/k_{\rm B} = 169~K$, but with a slightly higher prefactor value, $\tau_0^\text{sim}=6.8\times10^{-12}$~s. Hence, our data and simulations show a very good internal consistency, except for the $\tau_0$ value. This relatively minor difference,
we ascribe to anharmonicities in the anisotropy potential, not included in the simulations.

\subsection{Uncompensated moment and coupling of modes}
Contrary to expectations from simple theory, our simulations show coupling between the $\omega_\alpha$ and $\omega_\beta$ modes. This is most obvious in the power spectra in Fig.~\ref{fig:powersim}, where the $x$-component has a broad central peak. At lower temperatures, this signal consists of two peaks at $\pm \hbar \omega_\alpha$; the peaks gradually merge as the temperature is increased. By varying the simulation parameters, we found the coupling to be due to the uncompensated moment. At $\xi=1$, the modes seem to be completely uncoupled as expected. However, our simulations with $\xi = 1.010$ in our view represents closer the physical reality for a typical nanoparticle, and it is satisfying that even with this complication we are able to reproduce the experimental observations so accurately.


\subsection{Absence of elastic scattering}

As a final test of the rotor model, we return to the initial observation of the disappearance of the elastic 
neutron scattering signal when measuring with energy analysis on a triple-axis instrument.

\begin{figure}[]
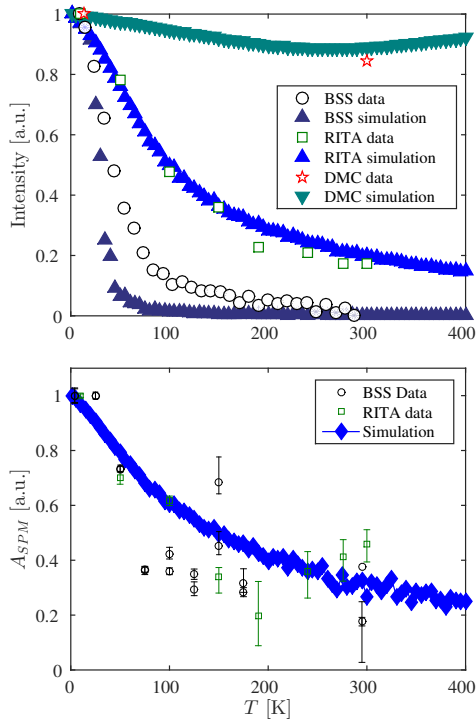

\includegraphics[width=0.35\textwidth]{RITA_and_BSS_and_sim_elastic_signal.eps}
\includegraphics[width=0.35\textwidth]{SPM_area.eps}
\caption{ (color online)
       Temperature dependence of the elastic part of the inelastic neutron scattering signal.
       (top) Background subtracted diffraction data from the backscattering spectrometer BSS, the triple-axis spectrometer RITA-2 and the diffractometer DMC.
       (bottom) The area of the quasielastic superparamagnetic signal as obtained from modeling the data from BSS and RITA-2.
\label{fig:Telastic}}
\end{figure}

The top panel of Fig.~\ref{fig:Telastic} shows the elastic magnetic signal when integrated over the energy resolution of BSS, RITA-II and DMC, respectively. With the resolution of BSS of $\sim 1$~$\mu$eV, there is almost no elastic signal at temperatures above $\sim 100$~K. At RITA, the elastic signal at 300~K has decreased to approximately 20\% of the low temperature signal, while within the DMC resolution, most of the signal is still seen as elastic at 300~K. 

 Overlayed on this, Fig.~\ref{fig:Telastic} shows the corresponding intensities from the numerical simulations,
scaled only by a common factor. The agreement between experiment and simulation is striking for the DMC and RITA-2 data, even though 
$A_{\rm SPM}$ was not used in the tuning of the simulations. This gives additional confidence
in the simulated model. 
In the backscattering data, the experimental value of the elastic signal is in general higher than the simulations.
This may well be a combination of two effects: a) the difference in $\tau_0$ value, b) a distribution of particle sizes, and therefore anisotropy barriers. Neither of these effects are included in the numerical simulations.

The bottom panel of Fig.~\ref{fig:Telastic} shows the total area of the fitted quasielastic magnetic signal, $A_{\rm SPM}$ as a function of temperature.  This value has been corrected for inelastic scattering and should thus be equal for both the BSS and RITA experiment, which is indeed the case within the uncertainties. Agreement with the simulations is also found in this case.


%
%

\subsection{Perspectives of the rotor mode}
Our analytical and numerical results provide a qualitative explanation 
of the experimentally observed increase of $\omega_\alpha$:
When thermal excitations cause a canting of the sublattice spins out of the basal plane,
a torque from the exchange interaction will appear.
This dominates the axial anisotropy, $B_2$,
and causes the spins to perform coherent precessions within the easy plane,
hence inducing periodic reversals of $\hat{S}^z$.
The frequency of the precession depends on the canting, and hence on temperature.
This displaces the neutron scattering signal from (quasi-) elastic to inelastic, in accordance with experimental observation.

%

We speculate that the rotor mode may be present in
other nanoparticles with a strong planar anisotropy,
{\em e.g.} NiO, although the temperature effect in this system was found to be less pronounced \cite{bahl06}.

\section{Conclusion}
We observe an increase with temperature of the mean frequency of the
uniform mode in weakly interacting antiferromagnetic 8~nm hematite particles.
Simultaneously, we observe a disappearance of the strictly elastic signal from static magnetic order.
Numerical Langevin simulations of a fairly detailed model is able to reproduce the features of the experimental data with sufficient accuracy. In combination with an analytical model, the simulation show that both observations
can be understood by the presence of a new fast dynamical mode
-- the ``rotor'' mode, which can be seen as a coherent superparamagnetic relaxation.
In contrast to other uniform modes in magnetic nanoparticles,
the rotor mode is driven by temperature-induced canting of the sublattice spins
and causes a periodic magnetisation reversal that transforms the spin system to become purely dynamical.
Our results are in good quantitative agreement with earlier neutron scattering and M\"ossbauer data.

We predict this mode to be of general relevance for antiferromagnetic nanoparticles
with strong easy-plane anisotropy.\\

\section*{Acknowledgments}
We thank F.~B\o dker for preparing the nanoparticles
and P.-A. Lindg\aa rd for stimulating discussions.
A large thank goes to S.~M\o rup and C.~Frandsen
for participating in the initial phases of this project.
This work was supported by the Danish Technical Research Council through
the Nanomagnetism framework program, and the
Danish Natural Science Research Council through DANSCATT.
This work is based on neutron scattering experiments performed at SINQ,
Paul Scherrer Institute, Switzerland, and at FRJ Research Center J\"ulich, Germany.

\end{document}